\pdfoutput=1
\documentclass[a4paper,11pt]{article}
\usepackage[utf8]{inputenc}

\usepackage[table]{xcolor}
\usepackage{jcappub}
\usepackage{graphicx}
\usepackage{psfrag,fancyhdr,epsfig}
\usepackage{hyperref}
\hypersetup{colorlinks,bookmarksopen,bookmarksnumbered,citecolor=brown,
linkcolor=blue,pdfstartview=FitH,urlcolor=blue}

\usepackage{amsmath}
\usepackage{color}
\usepackage{array}
\usepackage{graphicx}
\usepackage{color}
\usepackage{tensor}
\usepackage{xcolor}
\usepackage{orcidlink}

\newcommand{\be}{\begin{equation}}
\newcommand{\ee}{\end{equation}}
\newcommand{\bear}{\begin{array}}
\newcommand{\eear}{\end{array}}
\newcommand{\ba}{\begin{eqnarray}}
\newcommand{\ea}{\end{eqnarray}}

\def\a{\alpha}
\def\b{\beta}

\def\d{\delta}

\def\g{\gamma}

\def\l{\lambda}

\def\n{\nu}

\def\tns{\tensor}

\title{On the absence of ghosts in quadratic bigravity}


\author[a]{Ioannis D.~Gialamas\orcidlink{0000-0002-2957-5276}}
\author[b]{and Kyriakos Tamvakis\orcidlink{0009-0007-7953-9816}}

\emailAdd{ioannis.gialamas@kbfi.ee}
\emailAdd{tamvakis@uoi.gr}

\affiliation[a]{Laboratory of High Energy and Computational Physics, 
National Institute of Chemical Physics and Biophysics, R{\"a}vala pst.~10, Tallinn, 10143, Estonia}
\affiliation[b]{Physics Department, University of Ioannina, 45110, Ioannina, Greece}

\abstract{An extension of the bimetric theory of gravity is considered that includes quadratic Ricci curvature terms associated with each metric. The issue of the Boulware-Deser ghost is analyzed. The Hamiltonian constraint is derived and the existence of a secondary constraint is shown, proving that the theory is ghost-free.}

\begin{document}

\maketitle

\section{Introduction}

The theory of a massive spin-2 particle has a rather long history, starting with the linear theory of Fierz and Pauli in 1939~\cite{Fierz:1939ix}. Two decades later the problem of the so-called van Dam-Veltman-Zakharov discontinuity~\cite{vanDam:1970vg,Zakharov:1970cc} was cured by Vainshtein's nonlinear generalization~\cite{Vainshtein:1972sx}, while at around the same time Boulware and Deser showed that generic theories of massive gravity will necessarily have ghost modes~\cite{Boulware:1972yco}. The Boulware-Deser ghost can be avoided in a special class of massive gravity~\cite{deRham:2010kj,Hassan:2011tf,Hassan:2011vm} and bimetric gravity~\cite{Hassan:2011zd}  theories due to the existence of a pair of constraints that eliminate both the ghost field and its conjugate momentum~\cite{Hassan:2011hr,Hassan:2011ea}. Particularly, for the bimetric theory of gravity based on the introduction of a second dynamical metric tensor, these constraints were explicitly derived~\cite{Hassan:2011ea} and the absence of the Boulware-Deser ghost confirmed. As in the case of the standard treatment of gravitation in terms of general relativity, for the above bimetric theory it is assumed that its quantum aspects can be ignored at energies below the gravitational scale and, therefore, gravity can be treated classically. This is in contrast to the full quantum character of matter interactions which has to be taken into account. However, gravitating matter fields through their quantum interactions are expected to generate modifications to the gravitational action. Such a possible modification is the presence of quadratic terms of the curvature. A well known modification of the standard Einstein-Hilbert action of general relativity is the so called Starobinsky model~\cite{Starobinsky1980}, featuring a quadratic term of the Ricci scalar curvature, which has proven to be quite robust in its inflationary predictions~\cite{Planck:2018jri,BICEP:2021xfz}. Actually, a bimetric version of the Starobinsky model\footnote{ The influence of a quadratic curvature term in bigravity within the framework of metric-affine gravity has been investigated recently in~\cite{Gialamas:2023aim}. Additionally, for applications within the context of $F(R)$ massive gravity and bigravity, refer to~\cite{Kluson:2013yaa, Nojiri:2012re}.} has been proposed featuring quadratic terms $\sqrt{-g}R^2(g)$ and $\sqrt{-f}R^2(f)$ associated with each one of the two independent metric tensors $g_{ \mu\nu}$ and $f_{ \mu\nu}$~\cite{Gialamas:2023lxj}. This model describes successfully cosmic inflation~\cite{Gialamas:2023lxj} and could potentially account for the dark matter content of our Universe through the excitation of a massive spin-2 particle~\cite{Aoki:2014cla, Aoki:2016zgp, Babichev:2016hir, Babichev:2016bxi, Marzola:2017lbt, Manita:2022tkl, Kolb:2023dzp}. In the present article, we consider this model as a working example of quadratic bigravity and investigate the absence of the Boulware-Deser ghost. We perform an analysis in the framework of the Hamiltonian formulation, deriving the Hamiltonian constraint as well as the secondary constraint, sufficient to eliminate the Boulware-Deser ghost from the spectrum of the theory. 

The outline of the paper is as follows. In section~\ref{sec:2} we set up the theoretical framework of the quadratic bimetric gravity. Section~\ref{sec:3} discusses essential elements of the Hamiltonian formalism.  The Hamiltonian and the secondary constraints are presented in section~\ref{sec:4} and in section~\ref{sec:5} we compute the Poisson bracket,  that drives us to the elimination of the Boulware-Deser ghost. Finally, we summarize and conclude in section~\ref{sec:6}.

\section{Quadratic bimetric gravity}
\label{sec:2}
The standard bimetric action can be extended to include quadratic terms of the Ricci curvature scalars corresponding to each metric, namely
\be
{\cal{S}}=\int {\rm d}^4x\left\{m_g^2\sqrt{-g}\left(R(g)+\frac{\a_g}{2}R^2(g)\right)+m_f^2\sqrt{-f}\left(R(f)+\frac{\a_f}{2}R^2(f)\right)+2m_g^2m^2\sqrt{-g}V(\sqrt{\Delta})\right\}\,,{\label{ACT-0}}
\ee
where $g_{ \mu\nu}$ and $f_{ \mu\nu}$ are the two metric tensors and $\tns{\Delta}{^\mu_\n}=g^{ \mu\rho}f_{ \rho\nu}$.
 The physical Planck mass is given by $M_{\rm P}^2/2 =m_g^2+m_f^2$, while the parameter $m$ is redundant,
being an overall scale for the parameters $\b_n$. The parameters $\a_g$ and $\a_f$, are assumed to be constants with dimensions mass$^{-2}$. 
The special form of the potential $V$ is dictated by the consistency of the theory to be $V(\sqrt{\Delta})=\sum_{n=0}^{4}\beta_n\,e_n(\sqrt{\Delta})$ in terms of the parameters $\beta_n$ and the elementary symmetric polynomials of the square-root of the matrix $\Delta$, satisfying
\be 
e_n(X)=\frac{(-1)^{n+1}}{n}\sum_{k=0}^{n-1}(-1)^{k}Tr\left(X^{n-k}\right)e_k(X)\,,
\ee 
and starting with $e_0(\sqrt{\Delta})=1$. An equivalent form of ({\ref{ACT-0}}) can be written in terms of two scalars $\Omega$ and $\overline{\Omega}$ as
\begin{align}
{\cal{S}} =  \int{\rm d}^4x\bigg\{ & m_g^2\sqrt{-g}\,\Omega^2R(g)+m_f^2\sqrt{-f}\,\overline{\Omega}^2R(f)+2m^2m_g^2\sqrt{-g}\sum_{n=0}^{4}\beta_n e_n(\sqrt{\Delta}) \nonumber
\\
& -\frac{m_g^2}{2\a_g}\sqrt{-g}\left(\Omega^2-1\right)^2-\frac{m_f^2}{2\a_f}\sqrt{-f}\left(\overline{\Omega}^2-1\right)^2\bigg\}\,.
{\label{ACT-1}}
\end{align}
Performing a double Weyl-rescaling of the metrics\footnote{ Note that in the context of elementary symmetric polynomials, the relation $e_n(\sqrt{a\Delta}) = a^{n/2} e_n(\sqrt{\Delta})$ holds.}
\be 
g_{ \mu\nu}\rightarrow\,\Omega^{-2}g_{ \mu\nu} = e^{-\frac{\phi}{m_g\sqrt{3}}} g_{ \mu\nu}\,, \qquad f_{ \mu\nu}\rightarrow\,\overline{\Omega}^{-2}f_{ \mu\nu} = e^{-\frac{\overline{\phi}}{m_f\sqrt{3}}} f_{ \mu\nu}\,,
\label{eq:Weyl_res}
\ee
we transform the action to the Einstein frame and obtain
\begin{align}
\mathcal{S} =   \int {\rm d}^4x \bigg\{  &m_g^2 \sqrt{-g} R(g)+m_f^2\sqrt{-f} R(f)+2m^2m_g^2\sqrt{-g}\,\sum_{n=0}^{4}\tilde{\b}_n(\phi,\overline{\phi})e_n(\sqrt{\Delta})  \nonumber
\\ & -\frac{1}{2}\sqrt{-g}(\nabla\phi)^2 -\sqrt{-g}V_g(\phi)-\frac{1}{2}\sqrt{-f}(\nabla\overline{\phi})^2 -\sqrt{-f}V_f(\overline{\phi}) \bigg\}\,, 
\label{ACT-2}
\end{align}
with
\be
V_g(\phi) =  \frac{m_g^2}{2\a_g}\left(1-e^{-\frac{\phi}{m_g\sqrt{3}}}\right)^2\,, \qquad V_f(\overline{\phi}) = \frac{m_f^2}{2\a_f}\left(1-e^{-\frac{\overline{\phi}}{m_f\sqrt{3}}}\right)^2\,.
\ee
Note also that we have denoted $ (\nabla\phi)^2=g^{ \mu\nu}\nabla_{ \mu}\phi\nabla_{ \nu}\phi$, while $(\nabla\overline{\phi})^2=f^{ \mu\nu}\nabla_{ \mu}\overline{\phi}\nabla_{ \nu}\overline{\phi}$ . As anticipated, both scalar potentials align with the well-established Starobinsky form, as originally formulated in~\cite{Starobinsky1980}. Notably, these potentials exhibit the characteristic plateau, a  feature that manifests prominently when considering large field values. This plateau, reflects a stable and sustained region where the scalar field undergoes a slow-roll evolution, resulting in a prolonged period of inflationary expansion in absolute agreement with the latest observational results~\cite{Planck:2018jri,BICEP:2021xfz}. The parametric coefficients of the potential have now been promoted to the functions
\be \tilde{\beta}_n(\phi,\overline{\phi})=\beta_n\,e^{\frac{(n-4)\phi}{2m_g\sqrt{3}}}e^{-\frac{n\overline{\phi}}{2m_f\sqrt{3}}}\,.
\label{eq:betaphi}
\ee
As it is evident from~\eqref{ACT-2} the presence of each $R^2$ term introduces an extra propagating scalar degree of freedom (DOF). The model has been analyzed at the linear level in~\cite{Gialamas:2023lxj} and shown to describe a massless graviton, a massive spin-$2$ field and a pair of scalars without the presence of any ghost DOF. Furthermore, it was also shown in~\cite{Gialamas:2023lxj} that a combination of the scalars takes up the role of the inflaton with an inflationary potential analogous to that of the standard Starobinsky model~\cite{Starobinsky1980}.

Although we have not addressed the incorporation of matter coupled to the metric tensors directly, the gravitational scalars emerging in the Einstein frame can be viewed as matter. The introduction of matter and its permissible coupling with metric tensors has been extensively studied in the literature~\cite{Akrami:2014lja,Yamashita:2014fga,deRham:2014naa,deRham:2014fha,Gumrukcuoglu:2015nua,Melville:2015dba}. In our case, the scalar fields $\phi$ and $\overline{\phi}$ are coupled to different metrics, as indicated by the terms in the second line of eq.~\eqref{ACT-2}. In addition there is a mixed coupling term through the $\tilde{\beta}_n$ parameters, as defined in eq.~\eqref{eq:betaphi}. The proof of the absence of ghost fields that follows will also be a proof of the absence of ghosts for general matter coupled in this fashion to bigravity.

\section{Hamiltonian formalism}
\label{sec:3}
In order to analyze the physical content of the above action~\eqref{ACT-2} we consider the ADM Hamiltonian formulation~\cite{Arnowitt:1962hi} based on the decomposition of the metric in terms of the lapse and shift variables
\be
N = (-g^{00})^{-1/2}\,,\quad N_i =g_{0i}\,, \quad \g_{ij} = g_{ij}\,.
\label{LAPSE-1}
\ee
Similarly, for the second metric
\be
L = (-f^{00})^{-1/2}\,,\quad L_i =f_{0i}\,, \quad \tilde{f}_{ij} = f_{ij}\,.
\label{LAPSE-2}
\ee
Note that the raising and lowering of indices of $N^{i},\,L^{ i}$ is performed using the $3$-metrics $\g_{ij}$ and $\tilde{f}_{ij}$ according to $N_i = \g_{ij}N^j$ and $L_i = \tilde{f}_{ij}L^j$.
The momenta corresponding to $\gamma_{ij}$ can be obtained from the Einstein-Hilbert term of the Lagrangian to be 
\be \pi^{ij}=\frac{1}{m_g^2}\frac{\partial{\cal{L}}_{\rm EH}}{\partial\dot{\gamma}_{ij}}=N\sqrt{\gamma}\left(\tns{\Gamma}{^0_k_\ell}-g_{k\ell}\tns{\Gamma}{^0_m_n}g^{mn}\right)g^{ik}g^{j\ell}\,.\ee
Similarly, for the momenta $p^{ij}$ corresponding to the other metric $\tilde{f}_{ij}$. In terms of them the Einstein-Hilbert Lagrangian becomes
\be {\cal{L}}_{\rm EH}= m_g^2\pi^{ij}\dot{\gamma}_{ij}+m_f^2p^{ij}\dot{\tilde{f}}_{ij}+m_g^2R^{0}(\g)N+m_g^2R^{i}(\g)N_{ i}+m_f^2R^{0}(\tilde{f})L+m_f^2R^{i}(\tilde{f})L_i\,,\ee
where
\begin{subequations}
\begin{align}
& R^{0}(\g)=\sqrt{\gamma}\left[R(\gamma)+\gamma^{-1}\left(\frac{1}{2}\left(\tns{\pi}{^i_i}\right)^2-\pi^{i j} \pi_{i j}\right)\right]\,,\qquad R^{i}(\g)=2\sqrt{\gamma} \gamma_{i j} \nabla_{k}\left(\frac{\pi^{j k}}{\sqrt{\gamma}}\right)\,,
\\ & R^{0}(\tilde{f})=\sqrt{\tilde{f}}\left[R(\tilde{f})+\tilde{f}^{-1}\left(\frac{1}{2}\left(\tns{p}{^i_i}\right)^2-p^{i j} p_{i j}\right)\right]\,,\qquad R^{i}({\tilde{f}})=2 \sqrt{\tilde{f}} \tilde{f}_{i j} \nabla_{k}\left(\frac{p^{j k}}{\sqrt{\tilde{f}}}\right)\,.
\end{align}
\end{subequations}
$\gamma$ and $\tilde{f}$ are the determinants of the metrics $\gamma_{ij}$ and $\tilde{f}_{ij}$ and the dots denote derivatives with respect to time.
 The system encompasses 6 potentially DOFs, represented by the pair $\g_{ij}\,, \pi^{ij}$. Additionally, there are complementary DOFs arising from the pair $\tilde{f}_{ij}\,, p^{ij}$, culminating in a total of 12 DOFs. However, a crucial aspect unfolds in the subsequent section, where constraints are imposed, resulting in a reduction of the total DOFs to 7.
The remaining 7 DOFs correspond to the graviton (2 DOFs) and the massive spin-2 particle (5 DOFs).

The scalar part of the action~\eqref{ACT-2} can also be expressed in terms of the momenta 
\be
\varpi=\frac{\partial{\cal{L}}_{\phi}}{\partial(\partial_0\phi)}=\frac{\sqrt{\gamma}}{N}\left(\partial_0\phi-N^{i}\partial_i\phi\right)\,, \qquad\overline{\varpi}=\frac{\partial{\cal{L}}_{\overline{\phi}}}{\partial(\partial_0\overline{\phi})}=\frac{\sqrt{\tilde{f}}}{L}\left(\partial_0\overline{\phi}-L^{i}\partial_{i}\overline{\phi}\right)\,,
\ee
as
\be 
{\cal{L}}_{\phi,\,\overline{\phi}}=
\frac{N}{\sqrt{\gamma}}\left[\frac{\varpi^2}{2\gamma}-\frac{1}{2}\gamma^{ij}(\partial_i\phi)(\partial_{j}\phi)-V_g(\phi)\right]+\frac{L}{\sqrt{\tilde{f}}}\left[\frac{\overline{\varpi}^2}{2\tilde{f}}-\frac{1}{2}\tilde{f}^{ij}(\partial_i\overline{\phi})(\partial_{j}\overline{\phi})-V_f(\overline{\phi})\right]\,,
\ee  
while the corresponding scalar field Hamiltonians are given by
\begin{subequations}
\begin{align}
\mathcal{H}_\phi &= N\sqrt{\g} \left[ \frac{\varpi^2}{2\g} +\frac{1}{2} \g^{ij}(\partial_i \phi)(\partial_j \phi) +\frac{N^i}{N \sqrt{\g}}\varpi \partial_i \phi +V_g(\phi) \right]\,,
\\ \mathcal{H}_{\overline{\phi}} &= L\sqrt{\tilde{f}} \left[ \frac{\overline{\varpi}^2}{2\tilde{f}} +\frac{1}{2}\tilde{f}^{ij} (\partial_i \overline{\phi})(\partial_j \overline{\phi}) + \frac{L^i}{L \sqrt{\tilde{f}}}\overline{\varpi} \partial_i \overline{\phi} +V_f(\overline{\phi}) \right]\,.
\end{align}
\end{subequations}
For reasons that will be clarified in the upcoming section, we opt to decompose the aforementioned Hamiltonians as follows:
\be
\label{eq:H_dec}
\mathcal{H}_\phi = N \mathcal{H}^0_\phi +N^i {\mathcal{H}_\phi}_i\,, \qquad \mathcal{H}_{\overline{\phi}} = L  \mathcal{H}^0_{\overline{\phi}}  +L^i {\mathcal{H}_{\overline{\phi}}}_i\,,
\ee
with
\begin{subequations}
\begin{align}
 \mathcal{H}^0_\phi & =\sqrt{\g}\left(\frac{\varpi^2}{2\g} +\frac{1}{2} \g^{ij}(\partial_i \phi)(\partial_j \phi) +V_g(\phi)\right)\,, \qquad {\mathcal{H}_\phi}_i = \varpi \partial_i \phi\,, 
 \\
 \mathcal{H}^0_{\overline{\phi}} & =\sqrt{\tilde{f}}\left(\frac{\overline{\varpi}^2}{2\tilde{f}} +\frac{1}{2} \tilde{f}^{ij}(\partial_i \overline{\phi})(\partial_j \overline{\phi}) +V_f(\overline{\phi})\right)\,, \qquad {\mathcal{H}_{\overline{\phi}}}_i = \overline{\varpi} \partial_i \overline{\phi}\,.
\end{align}
\end{subequations}
It is evident from equation~\eqref{eq:H_dec} that the Hamiltonians exhibit linearity with respect to the lapse and shift variables. This characteristic holds significant importance for the subsequent analysis. Summarizing, the full Lagrangian takes up the form
\be
\mathcal{L} = \mathcal{L}_{\rm EH} + \varpi \dot{\phi} + \overline{\varpi} \dot{\overline{\phi}} -{\cal{H}}_{ \phi}^0N-{\mathcal{H}_\phi}_iN^i-{\cal{H}}^0_{\overline{\phi}}L-{\mathcal{H}_{\overline{\phi}}}_i L^i+2m^2m_g^2N\sqrt{\gamma}\sum_{n=0}^{4}\tilde{\beta}_n(\phi,\overline{\phi})e_n(\sqrt{\Delta}).
\label{LANG}
\ee

\section{Hamiltonian and secondary constraints}
\label{sec:4}
The variables $N,\,N_i,\,L,\,L_i$ appear in the Lagrangian~\eqref{LANG} without derivatives and, therefore, are non-dynamical, in contrast to the $\gamma_{ ij},\,\tilde{f}_{ ij},\,\phi$ and $\overline{\phi}$ variables which are dynamical along with their conjugate momenta. The equations of motion for the former make up the Hamiltonian constraint on the model. Nevertheless, since the potential depends on them in a highly non-linear fashion it is necessary to introduce new variables in terms of which the Hamiltonian constraint will become manifest. The four equations of motion for $N,\,N_i$ turn out to depend on three combinations, say $n_i$ (see~\cite{Hassan:2011tf}), providing the surplus equation as the Hamiltonian constraint. We, therefore, introduce new shift-like variables $n_i$ as
\be
\label{eq:nieq}
N^{i}-L^{i}=\left(L\tns{\d}{^i_j}+N \tns{D}{^i_j}\right)n^{ j}\,.\ee
The matrix $ \tns{D}{^i_j}$ is determined through the matrix relation
\be \sqrt{x}\,D=\sqrt{\left(\gamma^{-1}-Dnn^{T}D^{T}\right)\tilde{f}\,}\,, \label{eq:D}\ee
where $x=1-n^{i}\tilde{f}_{ ij}n^{j}$. In what follows,we will require only the condition~\eqref{eq:D},  and not an explicit solution for $\tns{D}{^i_j}$ in terms of $n^i$. However, it is crucial to emphasize that a solution does exist, as demonstrated in~\cite{Hassan:2011tf}. The resulting form of the Lagrangian in terms of the new variables is
\be
\mathcal{L} = m_g^2 \pi^{ij}\dot{\g}_{ij} + m_f^2 p^{ij}\dot{\tilde{f}}_{ij} + \varpi\dot{\phi} +\overline{\varpi} \dot{\overline{\phi}} - \mathcal{H}_0  + N \mathcal{C}\,,
{\label{LANG-1}}\ee
where $\mathcal{H}_0$ and $\mathcal{C}$ stand for 
\begin{align}
\mathcal{H}_{0} & =-L^{i}\left(m_g^2R_{i}(\g)+m_{f}^{2} R_{i}(\tilde{f}) + {\mathcal{H}_\phi}_i + {\mathcal{H}_{\overline{\phi}}}_i\right) \nonumber
\\ & \hspace{0.42cm}-L\left(m_{f}^{2} R^{0}(\tilde{f}) +m_g^2n^{i} R_{i}(\g) + {\mathcal{H}_{\overline{\phi}}}^0 + n^i {\mathcal{H}_\phi}_i +2 m^{2} m_g^2 \sqrt{\gamma} U\right)\,, \\
\mathcal{C} & =m_g^2R^{0}(\g)+m_g^2R_{i}(\g) \tns{D}{^i_j} n^{j} + \mathcal{H}_\phi^0 +{\mathcal{H}_\phi}_i\tns{D}{^i_j} n^{j}+2 m^{2} m_g^2 \sqrt{\gamma} \tilde{U}\,.
\end{align}
$U$ and $\tilde{U}$ are defined as
\begin{align} 
& U=\tilde{\beta}_{1}(\phi, \overline{\phi}) \sqrt{x}+\tilde{\beta}_{2}(\phi, \overline{\phi})\left(\sqrt{x}^{2} \tns{D}{^i_i}+n^{i} \tilde{f}_{i j} \tns{D}{^j_k} n^{k}\right) \nonumber
\\ & \hspace{0.7cm}+\tilde{\beta}_{3}(\phi, \overline{\phi})\left[\sqrt{x}\left(\tns{D}{^\ell_\ell} n^{i} \tilde{f}_{i j} \tns{D}{^j_k} n^{k}-\tns{D}{^i_k} n^{k} \tilde{f}_{i j} \tns{D}{^j_\ell} n^{\ell}\right)+\frac{1}{2} \sqrt{x}{ }^{3}\left(\tns{D}{^i_i} \tns{D}{^j_j}-\tns{D}{^i_j} \tns{D}{^j_i}\right)\right]  \nonumber
\\ &  \hspace{0.7cm}+\tilde{\beta}_{4}(\phi, \overline{\phi}) \sqrt{\tilde{f}}/\sqrt{\g}\,,
\\& \tilde{U}=\tilde{\beta}_{0}(\phi, \overline{\phi})+\tilde{\beta}_{1}(\phi, \overline{\phi}) \sqrt{x} \tns{D}{^i_i}+\frac{1}{2} \tilde{\beta}_{2}(\phi, \overline{\phi}) \sqrt{x}^{2}\left(\tns{D}{^i_i} \tns{D}{^j_j}-\tns{D}{^i_j} \tns{D}{^i_j}\right) \nonumber
\\& \hspace{0.7cm}+\frac{1}{6} \tilde{\beta}_{3}(\phi, \overline{\phi}) \sqrt{x}{ }^{3}\left(\tns{D}{^i_i} \tns{D}{^j_j} \tns{D}{^k_k}-3 \tns{D}{^i_i} \tns{D}{^j_k} \tns{D}{^k_j}+2 \tns{D}{^i_j} \tns{D}{^j_k} \tns{D}{^k_i}\right)\,.
\end{align}
Varying the Lagrangian~\eqref{LANG-1} with respect to the $n^{ i}$'s we obtain
 \be
\frac{\partial\mathcal{L}}{\partial n^k} = C_i \left[ L \tns{\d}{^i_k} +N \frac{\partial(\tns{D}{^i_j}n^j)}{\partial n^k}\right] = 0 \Rightarrow C_i = 0\,,
\label{eq:Cieq0}
\ee
with
\begin{align}
C_i  = & m_g^2 R_{i}(\g) - {\mathcal{H}_\phi}_i - 2m^2m_g^2 \sqrt{\g} \frac{n^\ell f_{\ell j}}{\sqrt{x}} \bigg[ \tilde{\b}_1(\phi,\overline{\phi}) \tns{\d}{^j_i} + \tilde{\b}_2(\phi,\overline{\phi}) \sqrt{x}(\tns{\d}{^j_i} \tns{D}{^m_m} -\tns{D}{^j_i}) \nonumber
\\ & +\tilde{\b}_3 (\phi,\overline{\phi}) {x} \left(\frac12 \tns{\d}{^j_i} (\tns{D}{^m_m} \tns{D}{^n_n} -\tns{D}{^m_n} \tns{D}{^n_m}) + \tns{D}{^j_m}\tns{D}{^m_i} -\tns{D}{^j_i} \tns{D}{^m_m}\right) \bigg]\,.
\label{eq:C_i}
\end{align}
This is independent of $N,\,L,\,N^{ i},\,L^{ i}$ and can in principle be solved to determine the $n^{ i}$'s as $n^{i}(\gamma,\,\pi,\,\tilde{f},\,\phi,\,\varpi,\,\partial\phi,\,\overline{\phi})$. 

Next, varying~\eqref{LANG-1} with respect to $N$ we obtain the Hamiltonian constraint
\be {\cal{C}}(n,\,\gamma,\,\pi,\,\tilde{f},\,\phi,\,\varpi,\,\partial\phi,\,\overline{\phi})\,=\,0\,. \label{eq:Ham_Con}\ee
 Now, using equations~\eqref{eq:C_i} and~\eqref{eq:Ham_Con} we can eliminate the lapse $N$ and the variables $n^i$ from the Lagrangian~\eqref{LANG-1}. The resulting expression remains linear in $L$ and $L^i$ (see~\cite{Hassan:2011zd}), so that they act as Lagrange multipliers, eliminating $1+3$ phase space DOFs.

 A secondary constraint arises when we demand that the Hamiltonian constraint~\eqref{eq:Ham_Con} is preserved with respect to time in the constraint surface ${\cal{C}}=0$. In the context of the Hamiltonian formulation, this implies that
\be
\frac{{\rm d}\mathcal{C}(x)}{{\rm d} t} = \{ \mathcal{C}(x), H\} = 0\,,\qquad \text{with}\qquad H = \int {\rm d}^3x' (\mathcal{H}_0(x') -N \mathcal{C}(x'))\,.
\ee
The Poisson bracket can be decomposed as
\be
\{ \mathcal{C}(x), H\} = \int {\rm d}^3x' \left\{ \mathcal{C}(x), \mathcal{H}_0(x')\right\} - \int {\rm d}^3x' N(x')\left\{ \mathcal{C}(x), \mathcal{C}(x')\right\} = 0\,.
\label{eq:CH0}
\ee
If $\left\{ \mathcal{C}(x), \mathcal{C}(x')\right\} \neq 0$, the condition~\eqref{eq:CH0} transforms into an equation for the lapse variable $N$ without imposing any additional constraints. The subsequent analysis will demonstrate that\footnote{The symbol $\approx$ refers to equalities that hold on a constraint surface, namely ${\cal{C}}=0$ in our case.} $\left\{ \mathcal{C}(x), \mathcal{C}(x')\right\}\, \approx\,0$ as in the case of pure massive gravity and standard bigravity~\cite{Hassan:2011ea}. Consequently, a secondary constraint will emerge as a result, namely 
\be
{\cal{C}}_{(2)}(x)=\int {\rm d}^3x'\left\{{\cal{C}}(x),{\cal{H}}_0(x')\right\}\,\approx\,0\,,
\ee
which will eliminate the canonical momentum conjugate to the ghost field. It is crucial to note that obtaining the secondary constraint relies on the Poisson bracket $\left\{ \mathcal{C}(x), \mathcal{H}_0(x')\right\}$ not vanishing identically. This has been shown in the case of standard bigravity~\cite{Hassan:2011ea} and it holds true in the present case as well. 

\section{Absence of the Boulware-Deser ghost}
\label{sec:5}

Before delving into confirming the absence of the Boulware-Deser ghost, let us count the DOFs of the model. The gravitational sector, represented by $\g_{ij}, \pi^{ij}$ and $\tilde{f}_{ij}, p^{ij}$, encompasses 12+12 phase space variables, while the scalar components, $\phi, \varpi$ and $\overline{\phi}, \overline{\varpi}$, contribute with extra 2+2 variables. In total, we have 28 variables.

Equation~\eqref{eq:Ham_Con} coupled with the equations $\d \mathcal{L}/\d L = 0$ and $\d \mathcal{L}/\d L^i = 0$, impose 5 constraints. The general coordinate invariance further eliminates 4 additional phase space DOFs through the gauge fixing in the process of the determination of the Lagrange multipliers $L$ and $L^i$.  Furthermore, once we have fixed $L$ and $L^i$ using the gauge freedom, the Lagrange multiplier $N$ is determined by the requirement that the secondary constraint ${\cal{C}}_{(2)}(x)$  is preserved under time evolution\footnote{It is important to mention that the consistency condition 
${\rm d}{\cal{C}}_{(2)}(x)/{\rm d}t=0$
determines $N$ as a function of the remaining variables and do not give rise to a tertiary constraint on the dynamical variables~\cite{Hassan:2011ea}.} (see appendix A of~\cite{Hassan:2018mbl} for further details). The shift variables $N^i$ are given by eq.~\eqref{eq:nieq}. This indicates that one can determine all Lagrange multipliers in bigravity by applying the consistency condition that ensures the preservation of ${\cal{C}}_{(2)}(x)$ over time, along with appropriate gauge fixing coordinate conditions. 

The last constraint, given by eq.~\eqref{eq:CH0} and to be verified in this section, serves to eliminate the Boulware-Deser ghost~\cite{Boulware:1972yco}. As indicated in Table~\ref{tab:my_label}, the 18 remaining variables correspond to 9 dynamical DOFs, namely the graviton (2 DOFs), the massive spin-2 particle (5 DOFs), the scalar $\phi$ (1 DOF), and the scalar $\overline{\phi}$ (1 DOF).
\begin{table}
    \centering
    \begin{tabular}{cc} 
     \hline    \rowcolor{gray!15} 
        $ \g_{ij}, \pi^{ij}, \tilde{f}_{ij}, p^{ij}, \phi, \varpi, \overline{\phi}, \overline{\varpi}$ & $28$ variables \\  \hline\\[-0.4cm]
        $\d \mathcal{L}/\d N = 0$~\eqref{eq:Ham_Con}  &  $1$ constraint \\
       $\d \mathcal{L}/\d L = 0$  &  $1$ constraint \\
        $\d \mathcal{L}/\d L^i = 0$  &  $3$ constraints \\
         general coordinate invariance & $4$ constraints \\
        secondary constraint~\eqref{eq:CH0} & $1$ constraint \\
        \hline
       \rowcolor{gray!15} Total & $\mathbf{18}$ variables \\ \hline 
    \end{tabular}
    \caption{The initial $28$ variables are reduced to $18$ after imposing the $10$ constraints. The $18$ remaining phase space variables correspond to $9$ dynamical DOFs, i.e. the graviton ($2$ DOFs), the massive spin-$2$ particle ($5$ DOFs), the scalar $\phi$ ($1$ DOF) and the scalar $\overline{\phi}$ ($1$ DOF). }
    \label{tab:my_label}
\end{table}

As mentioned earlier, in order to establish the existence of the secondary constraint, it is imperative to compute the Poisson bracket $\left\{ \mathcal{C}(x), \mathcal{C}(x')\right\}$. We proceed by separating the functions $\mathcal{C}$ and $\mathcal{C}_i$ into a part consisting of the terms present in the pure bigravity case, in which the $\phi,\,\overline{\phi}$ dependence enters only through the $\tilde{\beta}_n$'s, and a part that vanishes in the absence of the scalars. So,
\be
\mathcal{C} = \mathcal{C}_0 + \mathcal{C}_1 \quad \text{and} \quad \mathcal{C}_i = \mathcal{C}_i^{(0)} + \mathcal{C}_i^{(1)}\,,
\label{C_split}
\ee
with
\begin{subequations}
\begin{align}
\mathcal{C}_0 & =m_g^2{R^{0}}(\g)+m_g^2R_{i}(\g) \tns{D}{^i_j} n^{j} +2 m^{2} m_g^2 \sqrt{\gamma} \tilde{U} \,, 
\\
\mathcal{C}_1 & = - \mathcal{H}_\phi^0 - {\mathcal{H}_\phi}_i\tns{D}{^i_j} n^{j}\,,
\\
 \mathcal{C}_i^{(0)} & = m_g^2 R_{i}(\g)- 2m^2m_g^2 \sqrt{\g} \frac{n^\ell f_{\ell j}}{\sqrt{x}} \bigg[ \tilde{\b}_1(\phi,\overline{\phi}) \tns{\d}{^j_i} + \tilde{\b}_2(\phi,\overline{\phi}) \sqrt{x}(\tns{\d}{^j_i} \tns{D}{^m_m} -\tns{D}{^j_i}) \nonumber
\\ & \hspace{0.4cm} +\tilde{\b}_3 (\phi,\overline{\phi}) \sqrt{x}^2 \left(\frac12 \tns{\d}{^j_i} (\tns{D}{^m_m} \tns{D}{^n_n} -\tns{D}{^m_n} \tns{D}{^n_m}) + \tns{D}{^j_m}\tns{D}{^m_i} -\tns{D}{^j_i} \tns{D}{^m_m}\right) \bigg]\,, 
 \\
 \mathcal{C}_i^{(1)} & = - {\mathcal{H}_\phi}_i\,.
\end{align}
\end{subequations}
As a result the Poisson bracket $\left\{ \mathcal{C}(x), \mathcal{C}(x')\right\}$ is decomposed as
\be
\left\{ \mathcal{C}(x), \mathcal{C}(x')\right\} = \left\{ \mathcal{C}_0(x), \mathcal{C}_0(x')\right\} +\left\{ \mathcal{C}_0(x), \mathcal{C}_1(x')\right\} -\left\{ \mathcal{C}_0(x'), \mathcal{C}_1(x)\right\} +\left\{ \mathcal{C}_1(x), \mathcal{C}_1(x')\right\}\,.
\label{eq:C_dec}
\ee
We may simplify the notation denoting $\lambda^{i}=\tns{D}{^i_j}n^{ j}$.
In what follows we will compute the individual Poisson brackets given by~\eqref{eq:C_dec}, while in the appendix~\ref{appendix} we show all the relevant brackets in detail.  
\\[0.5cm]
I. \textbf{The bracket } $\mathbf{\left\{{\cal{C}}_0(x),\,{\cal{C}}_0(x')\right\}}$ \\[0.4cm]
The bracket $\left\{{\cal{C}}_0(x),{\cal{C}}_0(x')\right\}$ has been computed in~\cite{Hassan:2011ea} to be
\be 
\left\{{\cal{C}}_0(x),\,{\cal{C}}_0(x')\right\}=-\left[P^{i}(x)+P^{i}(x')\right]\partial_{ i}\delta^{(3)}(x-x')\,,
\label{eq:C0C0}
\ee
where 
\be
P^{i}(x)={\cal{C}}_0\lambda^{i}+\gamma^{i k}{\cal{C}}_{ k}^{(0)}+2{\cal{C}}_k^{(0)}\frac{\partial\lambda^{k}}{\partial\gamma_{ij}}\gamma_{ j\ell}\lambda^{ \ell}\,.
\label{eq:Pi}
\ee
In the case of standard bigravity, ${\cal{C}}_{i}^{(0)} = 0$, so the Poisson bracket is proportional to $\mathcal{C}_0$ which vanishes on the constraint surface.  In our case  ${\cal{C}}_{i}^{(0)} +{\cal{C}}_{i}^{(1)} = 0$ and the sum $\mathcal{C}_0 + \mathcal{C}_1$ vanishes on the constraint surface. 

To be precise, the Poisson bracket~\eqref{eq:C0C0} differs slightly from the one computed  in~\cite{Hassan:2011ea}. In our work, the term $\tilde{U}$ is not dependent on the parameters $\b_n$ but rather on the field-dependent parameters $\tilde{\b}_n(\phi,\overline{\phi})$.  Fortunately, all the Poisson brackets related to the scalar fields, vanish identically due to the absence of dependence on the conjugate momenta $\varpi$ and $\overline{\varpi}$.
\\[0.5cm]
 II. \textbf{The bracket } $\mathbf{\left\{ \mathcal{C}_0(x), \mathcal{C}_1(x')\right\} -\left\{ \mathcal{C}_0(x'), \mathcal{C}_1(x)\right\}}$  \\[0.4cm]
The mixed bracket is given by{\footnote{Quantities without a prime are intended to be calculated at the point $x$, whereas those with a prime should be calculated at $x'$.}
\begin{align}
\label{eq:c0c1}
\left\{ \mathcal{C}_0(x), \mathcal{C}_1(x')\right\} - \left\{ \mathcal{C}_0(x'), \mathcal{C}_1(x)\right\} = & -m_g^2 \left\{ R_{i}(\g), {\mathcal{H}_\phi^{0}}' \right\}\l^i + m_g^2 \left\{ {R_{i}(\g ')},\mathcal{H}_\phi^{0} \right\} {\l^{i}}' 
\\ & -m_g^2 \left\{ R_{i}(\g), {\l^{j}}' \right\} {{\mathcal{H}_\phi}_j}' \l^i  + m_g^2 \left\{ {R_{i}(\g')},\l^j \right\} {\mathcal{H}_\phi}_j {\l^{i}}'\,.  \nonumber
\end{align}
The two terms in the second line of eq.~\eqref{eq:c0c1} provide a term akin to the last term of eq.~\eqref{eq:Pi}  by replacing  $m_g^2 R_{i}(\g) \rightarrow - {\mathcal{H}_\phi}_i$ and $\tilde{U}\rightarrow 0$, that is 
\be 
-\left[\tilde{P}^{i}(x)+\tilde{P}(x')\right]\partial_{ i}\delta^{(3)}(x-x')\,, \quad{\text{where}}\quad \tilde{P}^{i}(x)=2{\cal{C}}_k^{(1)}\frac{\partial\lambda^{k}}{\partial\gamma_{ij}}\gamma_{ j\ell}\lambda^{ \ell}\,.
\label{eq:Pitil}
\ee
Thus, finally, we have
\begin{align}
\left\{ \mathcal{C}_0(x), \mathcal{C}_1(x')\right\} - \left\{ \mathcal{C}_0(x'), \mathcal{C}_1(x)\right\} =  -\bigg[&\tilde{P}^i(x)  + \frac{\varpi^2}{2\sqrt{\g}}\l^i -\sqrt{\g} \left(\frac12 (\nabla\phi)^2+V_g(\phi) \right)\l^i   \nonumber
\\ &+\sqrt{\g}\l^j\partial_j\phi\partial^i\phi + (x \rightarrow x') \bigg]\partial_{i} \d^{(3)}(x-x')\,.
\label{eq:C0C1}
\end{align}
\\[0.5cm]
\newpage
 III. \textbf{The bracket } $\mathbf{\left\{ \mathcal{C}_1(x), \mathcal{C}_1(x')\right\}}$  \\[0.4cm]
Next, we proceed to conclude the calculation by deriving the $\left\{{\cal{C}}_1(x),\,{\cal{C}}_1(x')\right\}$ bracket. We have
\be
\left\{ \mathcal{C}_1(x), \mathcal{C}_1(x')\right\} =  \left\{ \mathcal{H}_\phi^{0},{\mathcal{H}_\phi^{0}}'   \right\} + \left\{\mathcal{H}_\phi^{0}, {{\mathcal{H}_\phi}_j}'  \right\}{\l^j}'  - \left\{ {\mathcal{H}_\phi^{0}}', {\mathcal{H}_\phi}_j  \right\}\l^j + \left\{ {\mathcal{H}_\phi}_i,   {{\mathcal{H}_\phi}_j}'\right\}\l^i {\l^j}'\,,
\ee
which after substituting the relevant brackets from the appendix~\ref{appendix} becomes,
\be
\left\{ \mathcal{C}_1(x), \mathcal{C}_1(x')\right\} =\left({\mathcal{H}_\phi}^i +\frac{\varpi^2}{\sqrt{\g}}\l^i +\sqrt{\g}\l^j\partial_j\phi\partial^i\phi +\l^j {\mathcal{H}_\phi}_j \l^i + (x \rightarrow x') \right) \partial_i\d^{(3)}(x-x')\,.
\label{eq:C1C1}
\ee
Note that $\l^j {\mathcal{H}_\phi}_j = -\mathcal{C}_1 -\mathcal{H}_\phi^{0} $ and ${\mathcal{H}_\phi}^i = - \mathcal{C}_k^{(1)}\g^{k i}$.

To conclude, the total Poisson bracket, given by the sum of eqs.~\eqref{eq:C0C0}, \eqref{eq:C0C1} and~\eqref{eq:C1C1} is given by
\begin{align}
\left\{ \mathcal{C}(x), \mathcal{C}(x')\right\} =  \bigg[& -(\mathcal{C}_0 +\mathcal{C}_1)\l^i -(\mathcal{C}_k^{(0)}+\mathcal{C}_k^{(1)})\g^{k i} -2\left( \mathcal{C}_k^{(0)} +\mathcal{C}_k^{(1)}\right)\frac{\partial\lambda^{k}}{\partial\gamma_{ij}}\gamma_{ j\ell}\lambda^{ \ell} \nonumber
\\ & - \bigg( \frac{\varpi^2}{2\sqrt{\g}}\l^i -\sqrt{\g} \left(\frac12 (\partial\phi)^2+V_g(\phi) \right)\l^i +\sqrt{\g}\l^j\partial_j\phi\partial^i\phi \bigg) \nonumber
\\ & +\frac{\varpi^2}{\sqrt{\g}}\l^i +\sqrt{\g}\l^j\partial_j\phi\partial^i\phi - \mathcal{H}_\phi^{0}\l^ i  + (x \rightarrow x') \bigg]\partial_{i} \d^{(3)}(x-x')\,.
\end{align}
The second and the third terms of the first line vanish when we employ eqs.~\eqref{eq:Cieq0} and~\eqref{C_split}, whereas the second line is the opposite of the third line. So, the Poisson bracket takes the simple form:
\be
\left\{ \mathcal{C}(x), \mathcal{C}(x')\right\} =  -\left[ \left(\mathcal{C}_0(x) +\mathcal{C}_1(x)\right)\l^i   + \left(\mathcal{C}_0(x') +\mathcal{C}_1(x')\right){\l^i}' \right]\partial_{i} \d^{(3)}(x-x')\,.
\ee
Since, $ \mathcal{C}(x) = \mathcal{C}_0(x) +\mathcal{C}_1(x) = 0$ on the constraint surface, it is evident that $\left\{ \mathcal{C}(x), \mathcal{C}(x')\right\} \approx 0$. This result aligns with the fact that the Boulware-Deser ghost is indeed absent in this context.

\section{Conclusions}
\label{sec:6}
We considered bimetric gravity~\cite{Hassan:2011zd} extended to include quadratic Ricci curvature terms in the action, namely $\sqrt{-g} R^2(g)$ and $\sqrt{-f} R^2(f)$. This model is the bimetric extension of the Starobinsky model~\cite{Starobinsky1980} and it was originally proposed in~\cite{Gialamas:2023lxj}, where the inflationary dynamics are discussed. The field content of the model includes, apart from the standard massless graviton and the massive spin-$2$ field, two additional scalar DOFs, represented by two dynamical scalar fields. The model was analyzed in the framework of the Hamiltonian ADM formalism~\cite{Arnowitt:1962hi}, its physical content identified and the existing Hamiltonian constraint derived. Furthermore, a secondary constraint was proven to exist, as in the case of standard massive gravity and bigravity~\cite{Hassan:2011ea}. These constraints are sufficient to eliminate the Boulware-Deser field variable~\cite{Boulware:1972yco} as well as its conjugate momentum, proving that the model is ghost-free. In this way the initial 24 phase space variables are reduced to 14, corresponding to the aforementioned particles. This proof covers also models of bigravity coupled to matter in an analogous fashion.

\acknowledgments

The work of IDG was supported by the Estonian Research Council grants SJD18, MOBJD1202, RVTT3,  RVTT7, and by the CoE program TK202 ``Fundamental Universe''.

\appendix
\section{Relevant Poisson brackets} 
\label{appendix}

In this appendix we give the analytic expressions for the Poisson brackets used in the main sections. The basis of what follows are the canonical commutation relations:
\begin{align}
\label{eq:ap1}
\left\{\phi(x),\phi(x') \right\} = \left\{\varpi(x),\varpi(x') \right\} &= 0\,, \quad \left\{ \phi(x),\varpi(x') \right\} = \d^{(3)}(x-x')\,,
\\\left\{\g_{ij}(x), \g^{mn}(x') \right\} = \left\{ \pi_{ij}(x), \pi^{mn}(x') \right\}&=0\,, \quad\,\, \left\{\g_{ij}(x), \pi^{mn} \right\} = \frac12 \left( \tns{\d}{^m_i} \tns{\d}{^n_j} + \tns{\d}{^n_i} \tns{\d}{^m_j} \right) \d^{(3)}(x-x')\,.\nonumber
\end{align}
Now, using~\eqref{eq:ap1} we obtain\footnote{ To derive the results presented in section~\ref{sec:5} we have used the very useful identity of the $\d$-function, namely $$ A(x)B(x')\partial_x \d^{(3)}(x-x') -A(x')B(x)\partial_x'\d^{(3)}(x-x') = \left[A(x)B(x)+A(x')B(x') \right]\d^{(3)}(x-x')\,.$$ }:
\begin{align}
\left\{R^0(\g),F(\gamma')\right\} &=-\frac{1}{\sqrt{\gamma}}\left(\gamma_{mn}\pi^{ k}_{\,\,k}-2\pi_{mn}\right)\frac{\delta F(\gamma')}{\delta\gamma_{mn}}\,,
\\[0.1cm]
\left\{R_i(\g),F(\gamma')\right\} &= -\left(\gamma_{ im}{\nabla}_n+\gamma_{ in}\nabla_m\right)\frac{\delta F(\gamma')}{\delta\gamma_{ mn}}\,,
\\[0.1cm]
\left\{R^0(\g),R^0(\g')\right\} &= - \left[ R^i(\g) +R^i(\g')\right]\partial_i \d^{(3)}(x-x')\,,
\\[0.1cm]
\left\{R^0(\g),R_i(\g')\right\} &= - R^0(\g')\partial_i \d^{(3)}(x-x')\,,
\\[0.1cm]
\left\{R_i(\g),R_j(\g')\right\} &= - \left[ R_j(\g) \partial_i +R_i(\g') \partial_j \right]\d^{(3)}(x-x')\,,
\\[0.1cm]
\left\{R_i(\g), \mathcal{H}^0_\phi(x') \right\} &=\left[\frac{\varpi^2(x')}{2\sqrt{\g(x)}} -\sqrt{\g(x)} \left( \frac{\g'^{kj}}{2} \partial_k' \phi(x')  \partial_j' \phi(x') +V_g(\phi(x'))\right)\right]\partial_i \d^{(3)}(x-x') \nonumber
\\ & \hspace{0.5cm}+\frac{\sqrt{\g(x')}}{2} \partial_k' \phi(x')  \partial_j' \phi(x') \left(\g^k_i \partial^j +\g^j_i \partial^k \right) \d^{(3)}(x-x')\,,
\\[0.1cm]
\left\{\mathcal{H}^0_\phi(x),\mathcal{H}^0_\phi(x')\right\} &=\left[{\mathcal{H}_\phi}^i(x) + {\mathcal{H}_\phi}^i(x')\right]\partial_{ i}\delta^{(3)}(x-x')\,,
\\[0.1cm]
\left\{\mathcal{H}^0_\phi(x), {\mathcal{H}_\phi}_i(x') \right\} &= \left[\sqrt{\g(x)} \g^{kj}(x) \partial_i'\phi(x') \partial_k \phi(x) \partial_j +\frac{\varpi(x)\varpi(x')}{\sqrt{\g(x)}}\partial_i\right]\d^{(3)}(x-x') \nonumber
\\ & \hspace{0.5cm}+\partial_i'\phi(x') \frac{{\rm d} V_g(\phi(x))}{{\rm d} \phi(x)}\d^{(3)}(x-y)\,,
\\[0.1cm]
\left\{{\mathcal{H}_\phi}_i(x), {\mathcal{H}_\phi}_k(x') \right\} &= \left[\partial_k'\phi(x')\varpi(x) \partial_i   +\partial_i\phi(x)\varpi(x') \partial_k \right] \d^{(3)}(x-x')\,.
\end{align}
Poisson brackets of the form $\left\{R_i(\g),{\l^j}' \right\}$ and $\left\{ R_i(\g), \sqrt{\g'} \tilde{U}'\right\}$ have already been introduced in the main text and can be derived from equations~\eqref{eq:Pi} and/or~\eqref{eq:Pitil}. Note, also that many brackets, not listed above, (e.g. $\{ R^0(\g), \mathcal{H}^0_\phi(x') \}$) do not involve derivatives of the scalar $(\phi,\varpi)$ or gravitational $(\g_{mn}, \pi^{mn})$ pairs. As a result, these brackets are proportional to $\delta$-functions. When these terms and their conjugates (e.g. $-\{R^0(\g'), \mathcal{H}^0_\phi(x) \}$) are combined, their contribution is zero.

\vspace{1 cm}
\bibliography{bigravity_refs}{}

\end{document}